\documentclass[twocolumn,showpacs,preprintnumbers,amsmath,amssymb,prl]{revtex4}

\usepackage{graphicx}% Include figure files
\usepackage{dcolumn}% Align table columns on decimal point
\usepackage{bm}% bold math
\usepackage{color}

\begin{document}

\newcommand{\stgb}[5]{\ensuremath{\Sigma#1\left(#2#3#4\right)\theta=#5^\circ}}
\newcommand{\atgb}[8]{\ensuremath{\Sigma#1\left(#2#3#4\right)_1/\left(#5#6#7\right)_2\Phi=#8^\circ}}

\newcommand{\planef}[3]{\ensuremath{\left\{ #1#2#3 \right\} }}
\newcommand{\plane}[3]{\ensuremath{\left( #1#2#3 \right)}}
\newcommand{\dirf}[3]{\ensuremath{\left\langle #1#2#3 \right\rangle}}

\preprint{PRL}

\title{Energetic driving force for preferential binding of   \\ self-interstitial atoms to Fe grain boundaries over vacancies}% Force line breaks with \\

\author{M.A. Tschopp}
\email{mtschopp@cavs.msstate.edu}
\author{M.F. Horstemeyer}
\email{mfhorst@cavs.msstate.edu}
\affiliation{Center for Advanced Vehicular Systems \\ Mississippi State University, Starkville, MS 39759, USA}%
\author{F. Gao}%
\author{X. Sun}%
\author{M. Khaleel}%
 \affiliation{Fundamental and Computational Science Directorate \\ Pacific Northwest National Laboratory, Richland, WA 99352, USA}%

\date{\today}% It is always \today, today,
             %  but any date may be explicitly specified

\begin{abstract}

Molecular dynamics simulations of 50 Fe grain boundaries were used to understand their interaction with vacancies and self-interstitial atoms at all atomic positions within 20 \AA\ of the boundary, which is important for designing radiation-resistant polycrystalline materials.  Site-to-site variation within the boundary of both vacancy and self-interstitial formation energies is substantial, with the majority of sites having lower formation energies than in the bulk.  Comparing the vacancy and self-interstitial atom binding energies for each site shows that there is an energetic driving force for interstitials to preferentially bind to grain boundary sites over vacancies.  Furthermore, these results provide a valuable dataset for quantifying uncertainty bounds for various grain boundary types at the nanoscale, which can be propagated to higher scale simulations of microstructure evolution.           

\end{abstract}

% PACS, the Physics and Astronomy Classification Scheme.

%\keywords{Atomistic simulation; copper; aluminum; asymmetric tilt grain boundaries; energy; structural unit model; 9R phase; energy minimization; coherent twin; incoherent; inclination; grain boundary plane; partial dislocation} 
%Use showkeys class option if keyword display desired

\maketitle

Defects (vacancies and self-interstitial atoms) caused by radiation damage play an important role in the degradation of physical and mechanical properties of materials for nuclear applications \citep{Was2007}.  Future design of nuclear materials requires understanding the basic nature of radiation damage and its evolution in materials at the atomic level.  Radiation damage begins with the transfer of kinetic energy to a single primary knock-on atom, which then collides and displaces neighboring atoms, creating an atomic cascade of atoms that are displaced from their equilibrium lattice positions.  Following this initial event, equal numbers of vacancies and self-interstitial atoms are generated as atoms attempt to return to an equilibrium position within the lattice.  Since vacancies, self-interstitial atoms, and their clusters have such a profound effect on the physical and mechanical properties of these alloys, understanding how defects form, diffuse, and are absorbed by sinks is crucial to understanding radiation damage in fuel cladding materials.  Since fuel cladding materials are polycrystalline in nature, grain boundaries are a significant sink source for defects.  

Molecular dynamics (MD) simulations in single crystal, bicrystal, and nanocrystal structures have provided a fundamental understanding of cascade events and the production of defects in polycrystalline materials \citep{Phy1995,Per2001}.  Atomistic simulations have also been used to examine the interaction of defects with grain boundaries \citep{Sam2003,Bai2010,Gao2006,Kur2008,Gao2009,Ter2010}.  For example, MD studies in nanocrystalline metals have shown that grain boundaries act as a sink for interstitials \citep{Sam2003}. While these studies are instrumental to understanding the operating mechanisms induced by radiation, these types of studies have numerous grain boundaries with many confounding effects.  For instance, the microstructure will contain complex grain boundaries with different tilt and twist misorientations, triple junctions, and a distribution of grain sizes.  It is difficult to ascertain how GB character influences the interaction between defects and grain boundaries.   Therefore, bicrystal simulations have been used to explore the influence of grain boundary (GB) character on specific properties.  For instance, recently \citet{Bai2010} used MD simulations and temperature accelerated dynamics to show that grain boundaries can act as an effective sink for vacancies and interstitials through various mechanisms in Cu.  For one such mechanism, interstitials are loaded into the boundary, which then acts as a source, emitting interstitials to annihilate vacancies in the bulk.  While these studies are instrumental to understanding the long time scale mechanisms, it would be beneficial to use bicrystal MD simulations to also understand how GB character may impact the effectiveness of grain boundaries to act as sinks for defects.   Additionally, the resulting datasets can be used to provide the lower and upper bounds for uncertainty analysis in micro- and meso-scale simulations of microstructure evolution.   

In this paper, we have investigated the formation energies of vacancies and self-interstitial atoms (SIAs) as a function of location within/around the GB for 50 \dirf100 symmetric tilt grain boundaries in bcc Fe to energetically assess the GB sink strength.  Nanoscale simulations were required to capture the physics of vacancy and interstitial formation energies at the GB interface.  A parallel MD code, LAMMPS \citep{Pli1995}, was used to run all simulations in this work.  First, a GB database consisting of 50 \dirf100 symmetric tilt grain boundaries was generated using bicrystal simulation cells with 3D periodic boundary conditions \citep[cf.][]{Rit1996,Tsc2007c,Tsc2007d}.  A minimum distance of 12 nm between the two grain boundaries was used to eliminate any potential interaction between the two boundaries.  As with previous work \citep{Tsc2007c,Tsc2007d}, multiple initial configurations with different in-plane rigid body translations and an atom deletion criterion were used to properly access an optimal minimum energy GB structure via the Polak-Ribi\`{e}re conjugate gradient energy minimization.  For an initial generation of the structures, the updated version of the \citet{Men2003} interatomic potential for Fe was used.  This embedded-atom method \citep{Daw1984} potential has been shown to perform well in nanoscale simulations for nuclear applications \citep{Mal2010}.

\begin{figure}[tbh!]
  \centering
\begin{tabular}{c}
\includegraphics[width=3in,angle=0]{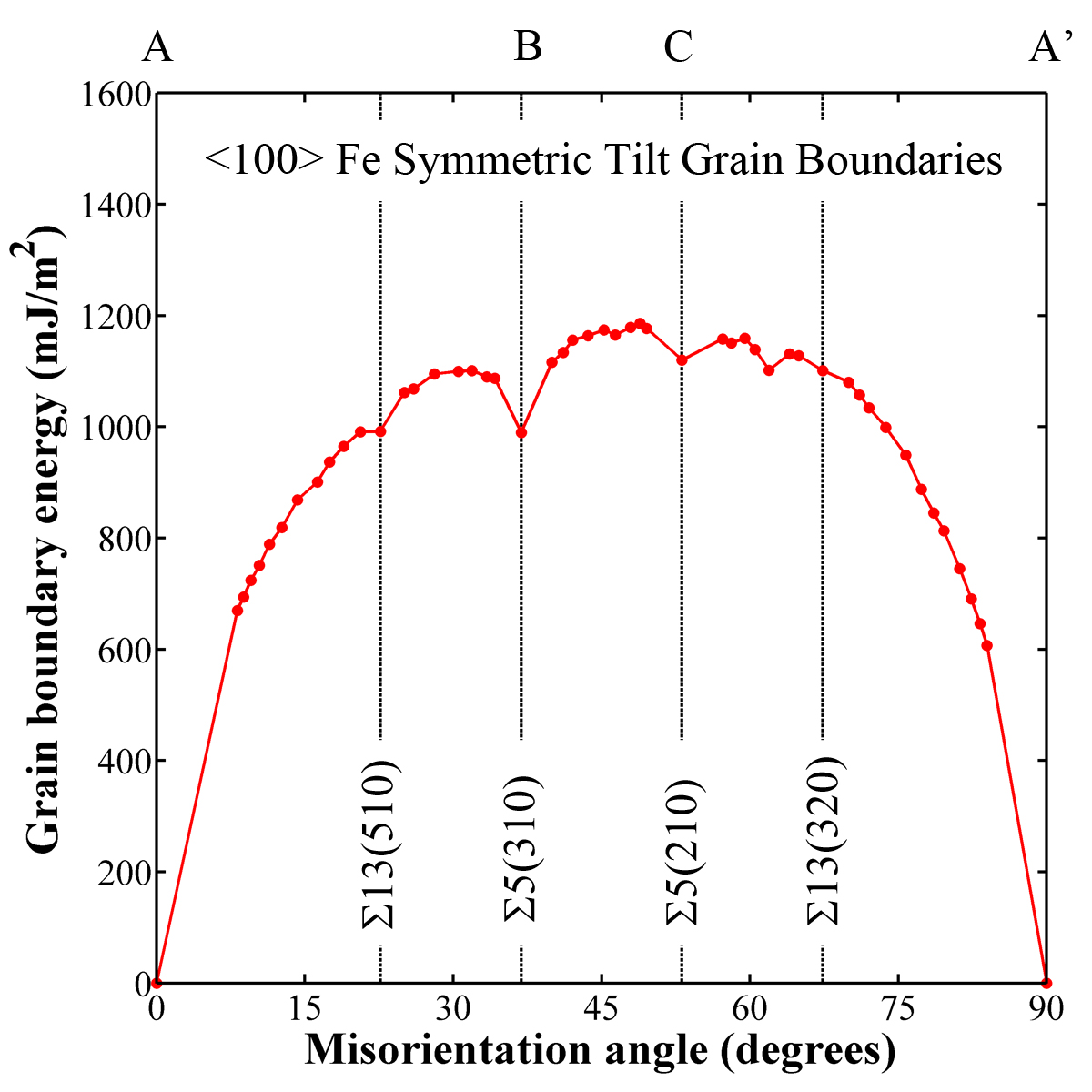} \\
\textbf{(a) } \\
\includegraphics[width=3in,angle=0]{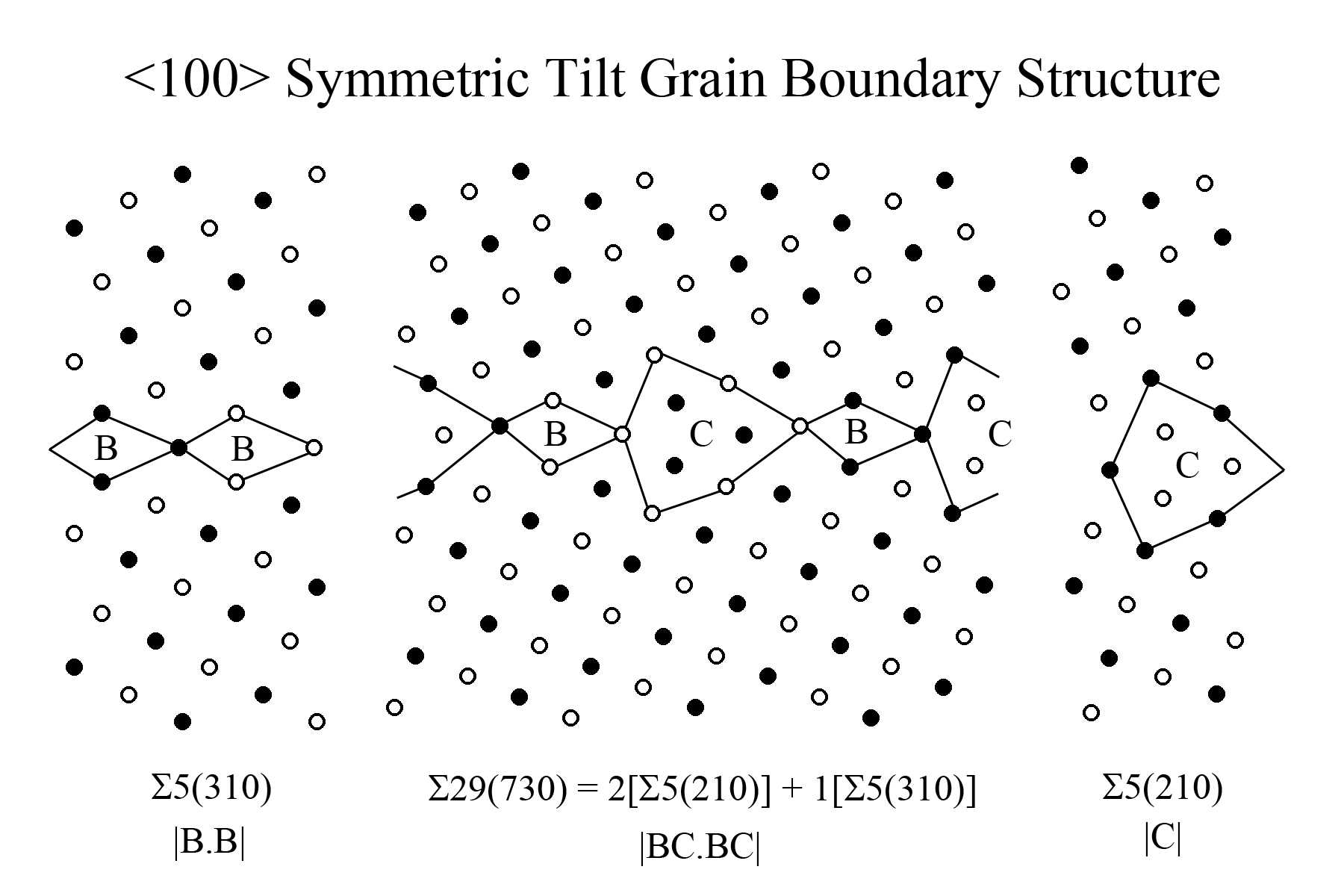} \\
\textbf{(b)}
\end{tabular}
\caption{(a) \dirf100 symmetric tilt grain boundary energy as a function of misorientation angle.  The low $\Sigma$ grain boundaries in each system are identified. (b) \dirf100 symmetric tilt grain boundary structures with structural units outlined for the $\Sigma5$\plane210, $\Sigma29$\plane730, and $\Sigma5$\plane310 STGBs.  Black and white denote atoms on different \planef100 planes. The different structural units are labeled B and C.}
  \label{fig:fig1}
\end{figure}

% Paragraph about Figure 1
A large number of grain boundaries were used to sample the range of GB structures and energies that might be observed in polycrystalline materials.  The \dirf100 symmetric tilt grain boundary (STGB) system chosen has several low order coincident site lattice (CSL) grain boundaries (the $\Sigma5$ and $\Sigma13$ boundaries) as well as both general high angle boundaries and low angle grain boundaries ($\leq15^\circ$).  Figure \ref{fig:fig1}a shows the GB energy as a function of misorientation angle for the \dirf100 symmetric tilt grain boundary system, similar to that found previously in Fe-Cr simulations \citep{Shi2008}.  The low-order CSL grain boundaries are also shown on this figure.  For the \dirf100 tilt axis, only minor cusps were observed in the energy relationship, most noticeably at the $\Sigma5$\plane310 boundary.  In addition to many general high angle boundaries, several low angle boundaries ($\leq15^\circ$) were also generated.  The range of GB energies sampled was $\approx{500}$ mJ/m$^2$.

%A large number of grain boundaries were used to sample the range of grain boundary structures and energies that might be observed in polycrystalline materials.  The \dirf100 symmetric tilt grain boundary system chosen has several low order coincident site lattice (CSL) grain boundaries (the $\Sigma5$ and $\Sigma13$ boundaries) as well as both general high angle boundaries and low angle grain boundaries ($\leq15^\circ$).  For electronic structures calculations, often only a few low order coincident site lattice grain boundaries are sampled due to the increased computational expense of higher $\Sigma$ boundaries (larger periodic distances and more atoms required).  However, using the embedded atom method potential within a molecular dynamics framework, 50 symmetric grain boundaries were easily generated and used for subsequent vacancy and interstitial formation energy calculations.  Figure \ref{fig:fig1}a shows the grain boundary energy as a function of misorientation angle for the \dirf100 symmetric tilt grain boundary system, similar to that found previously \citep{Shi2008}.  Both low order coincident site lattice (CSL) grain boundaries ($\Sigma5$ and $\Sigma13$) are also shown on this figure.  For the \dirf100 tilt axis, only minor cusps are observed in the energy relationship, most noticeably at the $\Sigma5$\plane310 boundary (990 mJ/m$^2$).  In addition to many high angle boundaries, several low angle boundaries ($\leq15^\circ$) were also generated.  The range of grain boundary energies sampled is approximately 500 mJ/m$^2$.

The GB structure plays an important role on the GB properties \citep{Mis2010}.  For low angle boundaries, the grain boundary is best represented by an array of discrete dislocations spaced a certain distance apart.  However, at higher misorientation angles the spacing between dislocations is small enough that dislocation cores overlap and dislocations rearrange to minimize the energy of the boundary.  The resulting GB structures are often characterized by structural units \citep{Sut1983}.  Grain boundaries with certain misorientation angles (and typically a low $\Sigma$ value) correspond to `favored' structural units, while all other boundaries are characterized by structural units from the two neighboring favored boundaries.  Figure \ref{fig:fig1}b shows an example for the \dirf100 STGB system, where the two $\Sigma5$ boundaries are favored STGBs and the $\Sigma29\plane730$ boundary is a combination of structural units from the two $\Sigma5$ boundaries.  The structural units for the $\Sigma5$\planef210 and $\Sigma5$\planef310 STGBs are labeled B and C, respectively, in a naming convention similar to that used for FCC metals \citep{Rit1996}.  Also notice that the ratio of structural units in the $\Sigma29$ GB can be determined by the crystallographic relationship of the two favored boundaries.  All boundaries between the favored structural units of the $\Sigma5$ boundaries and the `structural units' of the $0^\circ$ single crystals have a similar makeup.   

\begin{figure}[tbh!]
  \centering
\includegraphics[width=3.5in]{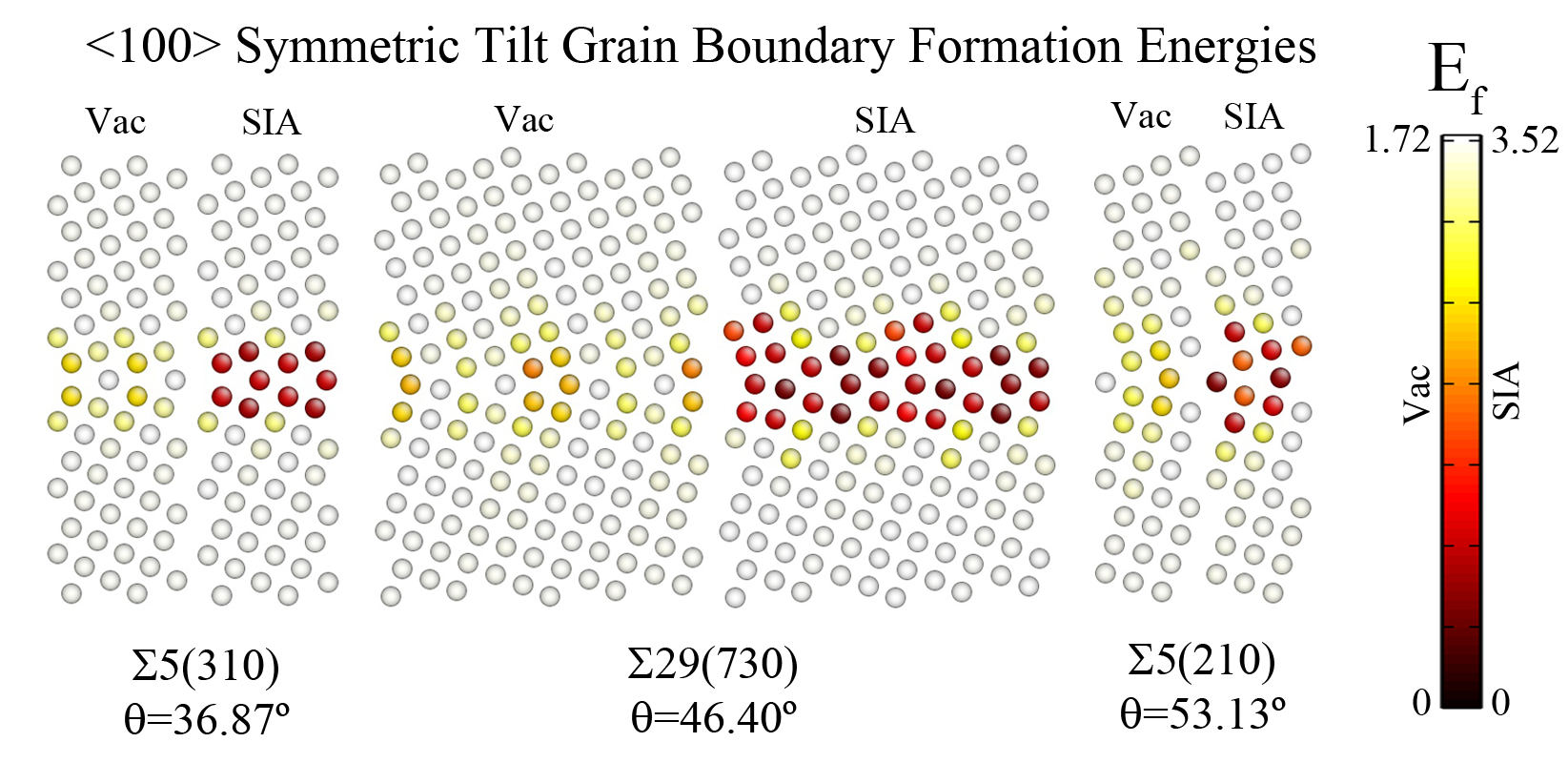} 
\caption{The vacancy and self-interstitial atom formation energies as a function of location for the $\Sigma5$\plane210, $\Sigma29$\plane730, and $\Sigma5$\plane310 STGBs.}
  \label{fig:fig2}
\end{figure}

\begin{figure}[bth!]
  \centering
\begin{tabular}{c}
\includegraphics[width=2.75in,angle=0]{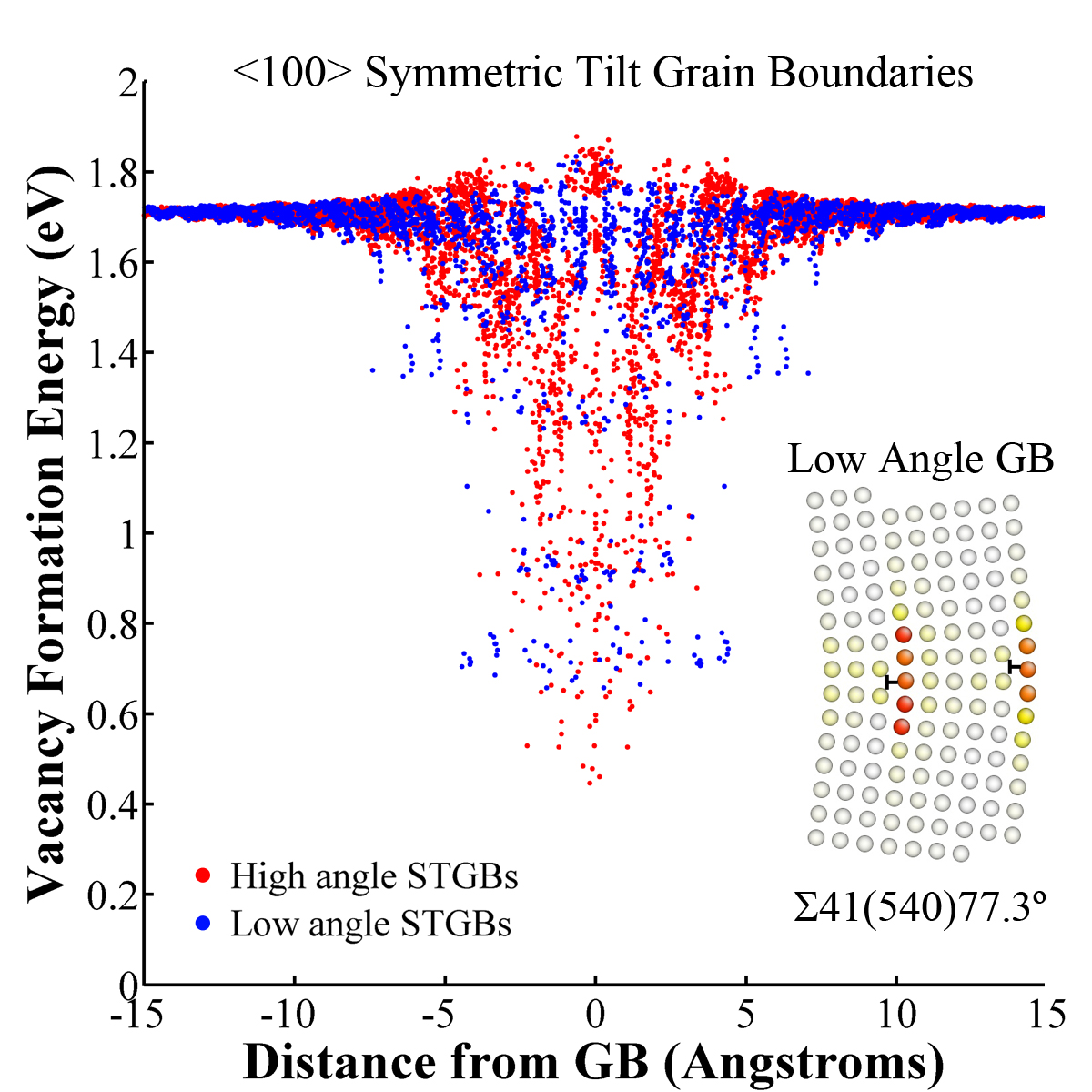} \\
\textbf{(a) } \\
\includegraphics[width=2.75in,angle=0]{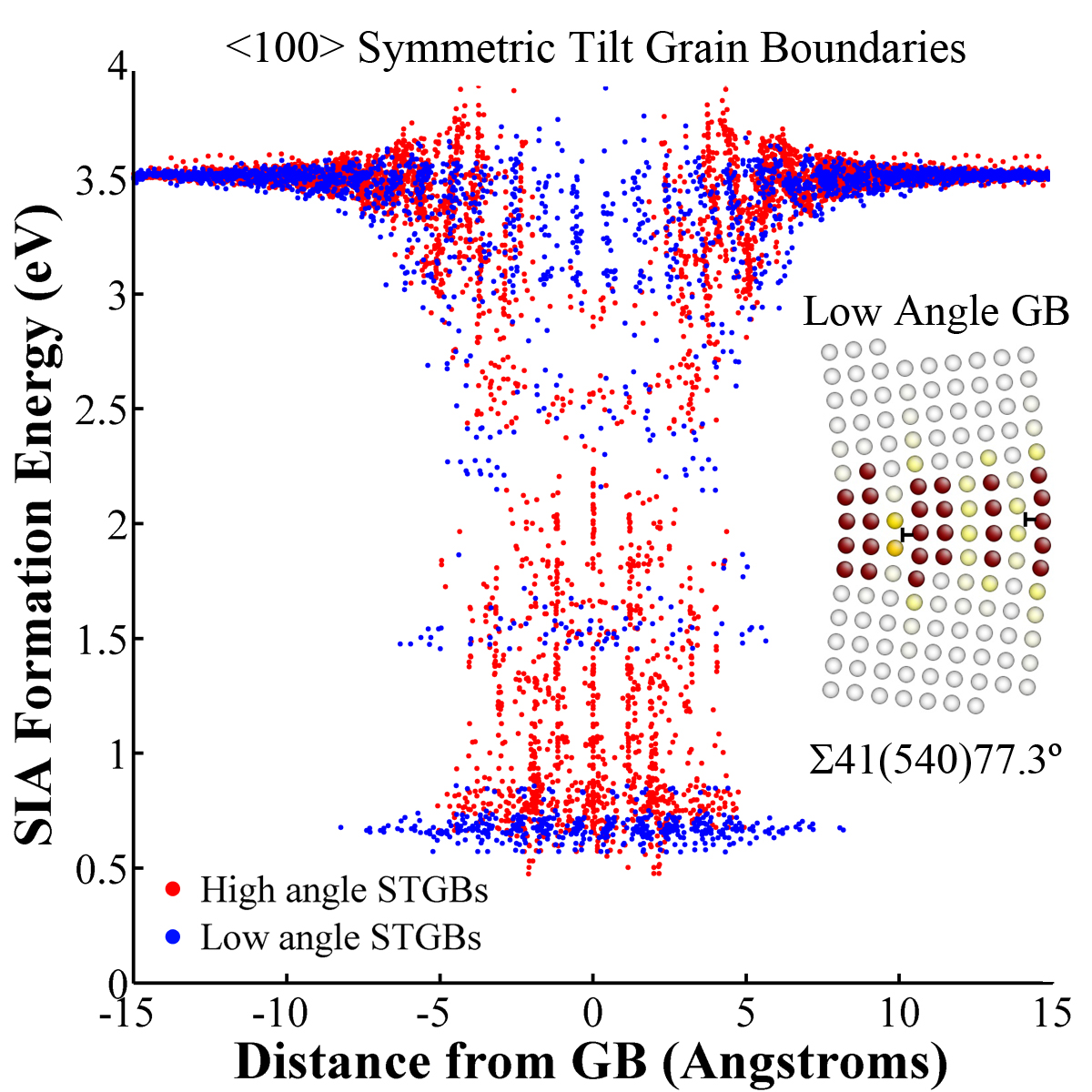} \\
\textbf{(b)}
\end{tabular}
\caption{ Evolution of (a) vacancy and (b) interstitial formation energies as a function of distance from the grain boundary for all 50 \dirf100 symmetric tilt grain boundaries (STGBs).  Low and high angle boundaries are colored differently.  The inset image is an example of a low angle boundary. }
  \label{fig:fig3}
\end{figure}

% Paragraph about Figure 2
The formation energies of vacancies and self-interstitial atoms at specific locations within/around the grain boundary were calculated for each grain boundary.  The approach here is similar to that used for modeling point defects in Cu \citep{Suz2003}.  First, all atoms within 20 \AA\ of each grain boundary were identified as potential sites.  Then, for vacancy formation energy simulations, an atom at a particular site $\alpha$ was removed and the simulation cell was relaxed through an energy minimization.    The vacancy formation energy for that particular site $\alpha$ was then calculated by,
\begin{equation}
  \label{eq:eq1}
	E_{f}^\alpha = E_{GB}^\alpha - E_{GB} + E_{coh}
\end{equation}
\noindent where $E_{coh}$ is the cohesive energy/atom of a perfect BCC lattice, and $E_{GB}^\alpha$ and $E_{GB}$ are the total energies of the simulation cell with and without the vacancy. On the other hand, for self-interstitial formation energy simulations, an atom was inserted approximately 0.5 \AA\  away at site $\alpha$, and the simulation cell was again relaxed through an energy minimization. The self-interstitial atom formation energy for site $\alpha$ was calculated by,
\begin{equation}
  \label{eq:eq2}
	E_{f}^\alpha = E_{GB}^\alpha - E_{GB} - E_{coh}
\end{equation}
The only difference between Eq.~\ref{eq:eq1} and Eq.~\ref{eq:eq2} is that $E_{coh}$ is subtracted, instead of added, to reflect the added energy due to placing that atom in an interstitial site.  This procedure was then performed for all 50 grain boundaries shown in Figure \ref{fig:fig1}(a) and the vacancy and interstitial formation energies were calculated as a function of location for each boundary.   

Figure \ref{fig:fig2} shows the vacancy and interstitial formation energies that correspond to atomic positions in the three GB structures shown in Fig.~\ref{fig:fig1}b.  The formation energy depicted for each location corresponds to a simulation.  In this graph, the colorbar is normalized such that the high value corresponds to the formation energy in the bulk, so that vacancy and interstitial energies can be easily compared.  For all three grain boundaries in Fig.~\ref{fig:fig2}, there are atoms lying along the GB plane that have higher vacancy formation energies similar to the bulk ($E_f=1.72$ eV).  However, there are several sites in each grain boundary that have lower vacancy formation energies than the bulk, suggesting an energetic driving force for vacancy diffusion from the single crystal to the grain boundary.  As the vacancy site is shifted away from the boundary, the formation energies approach that of the bulk.  For interstitials, most GB sites depicted here have a lower formation energy than in the bulk ($E_f=3.52$ eV).  AtomEye is used to visualize the simulation results \citep{Li2003}.

% Paragraph about Figure 3
The evolution of the (a) vacancy and (b) interstitial formation energies as a function of distance from the grain boundary is shown in Fig.~\ref{fig:fig3} for all 50 grain boundaries.  Interestingly, most of the formation energies that differ from the bulk occur within 5-8 \AA\  from the boundary center.  The majority of formation energies within this GB-affected region are lower than in the bulk.  However, for both vacancy and self-interstitial atoms, there are locations with higher formation energies than in the bulk, with the highest vacancy formation energies near the boundary center.  Additionally, this plot separates the high angle boundaries from the low angle boundaries, which is motivated by previous studies \citep[\textit{e.g.},][]{Gao2009}.  For interstitial formation energies, most high angle grain boundaries have much lower formation energies than that in the bulk.  The majority of interstitial formation energies near bulk values within 5 \AA\  of the boundary occur in the single crystal regions between dislocations for low angle boundaries.  For example, the inset image shows an example of a low angle boundary.  The visible dislocations in the low angle boundary have a local effect on formation energies and increasing the dislocation spacing (\textit{i.e.}, lower misorientation angle) merely results in shifting these localized regions further apart.  For low angle boundaries, the formation energies trend to that of an isolated dislocation in a single crystal.

\begin{figure}[tbh!]
  \centering
\includegraphics[width=2.75in,angle=0]{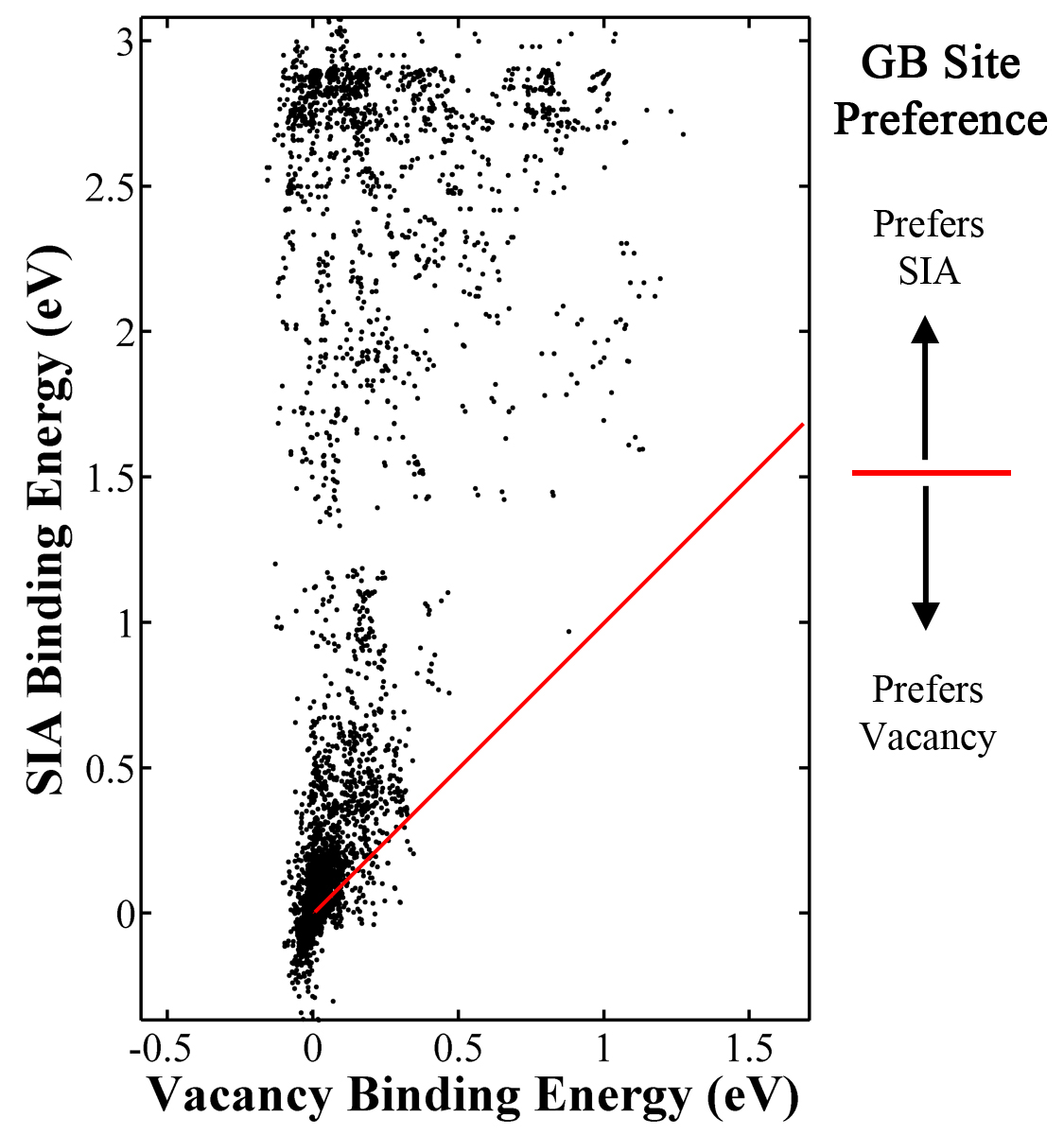} 
\caption{Grain boundary binding energies for vacancies and self-interstitial atoms (SIA) for all sampled sites in this study.  The line delineates sites where interstitial binding energy is greater than vacancy binding energies (above the line).}
  \label{fig:fig4}
\end{figure}

% Paragraph about Figure 4
The grain boundary binding energy for vacancies and self-interstitial atoms are plotted against each other for each site in Fig.~\ref{fig:fig4}.  The GB binding energy for a particular site $\alpha$ is calculated by subtracting the formation energy from the bulk formation energy, $E_{binding}^\alpha=E_{f,bulk}-E_f^\alpha$.  The line delineates sites where interstitial binding energy is greater than vacancy binding energies (above the line).  The large amount of binding energies above this line indicates that the system energy is decreased more through interstitials occupying GB sites, rather than vacancies.  Hence, this letter shows that there is an energetic driving force for interstitials to segregate to GB sites over vacancies.  This study also supports the interstitial loading-unloading mechanism \citep{Bai2010} by showing that it is energetically favorable for interstitials to initially `load' grain boundaries for a wide range of grain boundary types.  This is significant for nuclear applications where radiation damage generates lattice defects and grain boundaries act as sinks for both vacancies and interstitial atoms.  

In summary, this letter shows that self-interstitial atoms have a larger energetic driving force for binding to grain boundaries than vacancies, based purely on formation energies of these point defects in thousands of GB sites.  Fifty \dirf100 symmetric tilt grain boundaries were utilized here to sample some of the variability in GB degrees of freedom that is observed in experimental polycrystalline materials, including general low and high angle grain boundaries as well as some low-$\Sigma$ boundaries.  By iteratively calculating the formation energies of both vacancies and self-interstitial atoms at every site within 20 \AA\  of the boundary, this study was able to methodically examine the binding energy associated with GB segregation of point defects and calculate the energetic driving force for point defects to grain boundaries in Fe.

\begin{acknowledgments}

This work was funded by the U.S. Department of Energy's Nuclear Energy Advanced Modeling and Simulation (NEAMS) program at Pacific Northwest National Laboratory.  PNNL is operated by Battelle Memorial Institute for the U.S. Department of Energy under contract No. DE-AC05-76RL01830.

\end{acknowledgments}

%\section*{References}
%\bibliography{references}
%\bibliographystyle{unsrtnat}

\newpage
\end{document}